\documentclass{pasj00}

\SetRunningHead{H. Seta et al.}{Suzaku X-Ray Observations of Fornax A}
\Received{2013 May 4}
\Accepted{2013 July 1}
\begin{document}

\title{Suzaku Detection of Thermal X-Ray Emission\\
 Associated with the Western Radio Lobe of Fornax~A}

\author{%
   Hiromi \textsc{Seta}\altaffilmark{1},
   Makoto S. \textsc{Tashiro}\altaffilmark{2}, and
   Susumu \textsc{Inoue}\altaffilmark{3}
}

\altaffiltext{1}{Research Center for Measurement in Advanced Science, Faculty of Science, Rikkyo University 3-34-1 Nishi-Ikebukuro,Toshima-ku, Tokyo Japan 171-8501}
\email{seta@rikkyo.ac.jp}
\altaffiltext{2}{Department of Physics, Saitama University, 255 Shimo-Okubo, Sakura-ku, Saitama Japan 338-8570}
\altaffiltext{3}{Max-Planck-Institut f\"ur Kernphysik, Saupfercheckweg 1, 69117 Heidelberg Germany}

\KeyWords{%
galaxies: individual (Fornax~A) -- 
galaxies: magnetic fields --
radiation mechanisms: non-thermal -- 
radio continuum: galaxies -- 
X-rays: galaxies
}

\maketitle

\begin{abstract}
 We present the results of X-ray mapping observations of the western radio lobe of the
 Fornax A galaxy, using the X-ray Imaging Spectrometer (XIS) onboard the Suzaku
 satellite with a total exposure time of 327~ks. The purpose of this study is to
 investigate the nature and spatial extent of the diffuse thermal emission around the
 lobe by exploiting the low and stable background of the XIS. The diffuse thermal
 emission had been consistently reported in all previous studies of this region, but its
 physical nature and relation to the radio lobe had not been examined in detail.  Using
 the data set covering the entire western lobe and the central galaxy NGC\,1316, as well
 as comparison sets in the vicinity, we find convincingly the presence of thermal
 plasma emission with a temperature of $\sim$1~keV in excess of conceivable background and
 contaminating emission (cosmic X-ray background, Galactic halo, intra-cluster gas of
 Fornax, interstellar gas of NGC\,1316, and the ensemble of point-like sources). Its
 surface brightness is consistent with having a spherical distribution peaking at the
 center of the western lobe with a projected radius of $\sim$12\arcmin. If the volume
 filling factor of the thermal gas is assumed to be unity, its estimated total mass
 amounts to $\sim10^{10}$\,$M_{\odot}$, which would be $\sim$10$^{2}$ times that of the
 central black hole and comparable to that of the current gas mass of the host galaxy.
 Its energy density is comparable to or larger than those in the magnetic field and
 non-thermal electrons responsible for the observed radio and X-ray emission.
\end{abstract}

\section{Introduction}\label{s1}
Radio galaxies exhibit powerful outflows of magnetized plasma that emanate as collimated
jets from their nuclei and end in extended structures known as lobes.  The jets and
lobes are most prominently observed in the radio band through the synchrotron emission
of relativistic electrons that are accelerated at different locations in the outflow.
Although the jets are believed to be powered by the super-massive black hole residing in
the nucleus of the host galaxy, the basic physical mechanisms that drive their formation
and evolution are still poorly understood.  The lobes are thought to result from the
interaction of the outflow with the external intergalactic or intra-cluster medium (ICM),
whereby the kinetic energy of the jets is dissipated and their momentum isotropized
(e.g. \cite{begelman84}).  It has become increasingly apparent in recent years that
these lobes also constitute the interface through which jets strongly influence the
evolution of their host galaxies, groups, and/or clusters by heating and redistributing
the ambient gas in different ways.  However, the actual physical processes responsible
for such effects are currently under intensive debate \citep{mcnamara07,mcnamara12}.
Detailed studies of the composition and energy balance of different components inside
the lobes of radio galaxies should therefore provide valuable insight into the physics
of not only the jets themselves, but also of feedback effects that are crucial for the
evolution of large-scale structure in the Universe.

Fornax A (NGC\,1316; hereafter For A) is one of the closest and brightest radio galaxies, which is known
to have a distinctive, two-sided lobe structure. It is the brightest in the GHz band
\citep{eckers83}, and was the first to be reported to have diffuse, non-thermal X-ray
emission, which is generated by inverse Compton up-scattering of ambient radiation
(primarily the cosmic microwave background; \cite{feigelson95,kaneda95}).  Since then, many
other radio galaxies have been observed with X-ray emission from large-scale jets
\citep{harris06}. Although For A can be considered to be an archetypal object for
non-thermal X-ray emitting lobes, it is somewhat atypical in the internal energy
balance. The ratio of the energy densities in non-thermal electrons and the magnetic
field, as estimated from the relative strengths of the inverse Compton and synchrotron
emission, is exceptionally low compared to the lobes of other objects \citep{isobe09}.
This may be somehow related to the unique evolutionary state of For A; the unusually low
X-ray luminosity of the nucleus of its host galaxy NGC\,1316
(5$\times$10$^{39}$~erg~s$^{-1}$ in 0.3--8.0~keV; \cite{kim03}) suggests that the
central black hole ceased its activity some 0.1~Gyr ago from \cite{iyomoto98} (See also
\cite{lanz10}).
Emission
from such relic radio lobes whose power supply has been switched off are expected to
fade away not long afterwards, as the residual non-thermal electrons will only cool via
radiative and Coulomb losses.

For most radio galaxies, observations up to now have indicated that their lobes do not
contain a significant amount of thermal material \citep{mcnamara07}. For~A stands out as
a unique sample in its apparent existence of diffuse thermal X-ray emission around its
lobes.  In addition to the diffuse non-thermal X-ray emission of  inverse Compton
origin, all previous X-ray studies have consistently suggested that thermal emission is
also present in the spectra for both lobes on either side of the nucleus
\citep{feigelson95,kaneda95,tashiro01,isobe06,tashiro09}. This thermal emission had been
ascribed to hot gas of either the Fornax cluster or around the NGC\,1316 galaxy.
However, the alternative possibility that the emission actually originates from within
the radio lobes had not been seriously investigated, due to the lack of
spatially-resolved X-ray spectroscopy with low-background covering the entire lobe
structure.  Of further interest is the recent, possible detection of diffuse thermal
emission in the giant, outer lobes of the Centaurus A (hereafter Cen A) radio galaxy
using the Suzaku satellite \citep{stawarz13}.

\medskip

In this paper, we revisit the issue of diffuse thermal emission around the For A radio
lobe.  We present the results of the first X-ray mapping observations to cover the
entire western lobe of For~A, with an extent of 20\arcmin\ in diameter, as well as its
surroundings to examine the presence of the diffuse thermal emission and its association
with the radio lobe.

The plan of this paper is as follows. In \S~\ref{s2}, we describe the Suzaku mapping
observations of the western lobe and their data reduction. In \S~\ref{s3}, we present
the X-ray image (\S~\ref{s3-1}) and spectra, for which we evaluate the contributions of
various conceivable sources of background and contaminating emission (\S~\ref{s3-2} and
\ref{s3-3}). We conclude that the diffuse thermal emission does originate from within
the lobe and investigate its spatial distribution (\S~\ref{s3-4}). In \S~\ref{s4}, we
derive the total mass and energy of the diffuse gas and compare it with other forms of
 energy, as well as with the results for the Cen A outer lobe
\citep{stawarz13,osullivan13}. We summarize the main findings of this paper in
\S~\ref{s5}.

Throughout this paper, we assume a distance of 18.6~Mpc and a redshift of
4.65$\times$10$^{-3}$ for For A \citep{madore99}. We evaluate the observed surface
brightness of the diffuse emission in units of erg~s$^{-1}$~cm$^{-2}$~str$^{-1}$,
without correcting for Galactic absorption and assuming a uniform projected distribution,
unless otherwise noted. Fiducial energy bands of 0.5--2.0~keV and 2.0--10.0~keV are chosen
as typical ranges for the thermal and non-thermal components, respectively.

\section{Observations \& Data Reduction}\label{s2}
We performed six mapping observations of the western lobe of For~A using the Suzaku
satellite \citep{mitsuda07}. Suzaku has two kinds of operational instruments; the X-ray
Imaging Spectrometer (XIS; \cite{koyama07}) and the non-imaging Hard X-ray Detector
\citep{kokubun07,takahashi07}. We concentrate on the XIS data in this paper, 
because we focus on the spatial distribution of the thermal emission around the lobe.

\begin{figure}
 \begin{center}
  \FigureFile(85mm,85mm){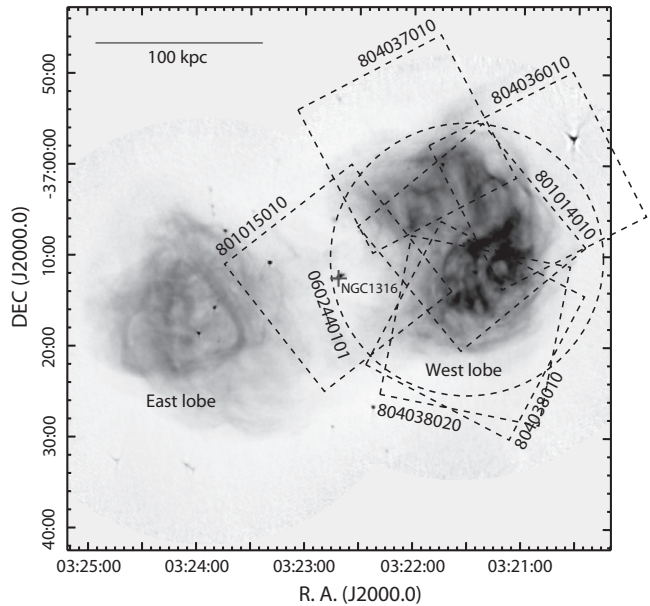}
 \end{center}
 \caption{Mapping layout of the Suzaku observations overlaid on a 1.5~GHz radio intensity map
 in gray scale \citep{fomalont89}. The Suzaku XIS and XMM-Newton MOS fields of view
 are shown as dashed squares and circles, respectively, with the sequence ID labels as in
 table~\ref{t1}. The position of the nucleus of NGC\,1316 is shown with a
 cross.}
 \label{f1}
\end{figure}

The mapping layout is shown in figure~\ref{f1}, while the observation details are
summarized in table~\ref{t1}. Two observations were made in 2006 and the remaining four
in 2009. The data for the central part of the western lobe in the earlier observation
(sequence ID 801014010 in figure~\ref{f1}) has already been presented in
\citet{tashiro09}. In this paper, we assembled the additional data sets taken in the flanking
fields to complete the coverage of the western lobe and the central galaxy (NGC\,1316)
of For~A with a total exposure time of 327~ks.

\begin{table*}
 \begin{center}
  \caption{Suzaku and XMM-Newton data sets for For~A.}
  \label{t1}
  \begin{tabular}{llcccc}
   \hline
   \hline
   Observatory  & Sequence ID  & \multicolumn{2}{c}{Aiming position (J2000.0)}   & Observation & $t_{\mathrm{exp}}$\footnotemark[$*$]\\
                &              & R.\,A. & DEC                & start date  & (ks)\\
   \hline
   Suzaku \dotfill     & 801015010  & \timeform{03h22m40.4s} & \timeform{-37D12'10''} & 2006-12-22 & 86.7\\
                       & 801014010  & \timeform{03h21m40.4s} & \timeform{-37D09'52''} & 2006-12-23 & 42.9\\
                       & 804036010  & \timeform{03h20m53.3s} & \timeform{-37D02'03''} & 2009-06-08 & 54.8\\
                       & 804037010  & \timeform{03h22m05.5s} & \timeform{-37D58'03''} & 2009-06-09 & 55.5\\
                       & 804038010  & \timeform{03h21m25.9s} & \timeform{-37D18'39''} & 2009-06-30 & 47.0\\
                       & 804038020  & \timeform{03h21m25.7s} & \timeform{-37D18'40''} & 2009-08-02 & 39.6\\
   XMM-Newton \dotfill & 0602440101 & \timeform{03h21m29.3s} & \timeform{-37D11'30''} & 2009-06-25 & 59.5\\
   \hline
   \multicolumn{6}{@{}l@{}}{\hbox to 0pt{\parbox{130mm}{\footnotesize
   \par\noindent
   \footnotemark[$*$] The mean exposure time of the three XIS sensors for Suzaku and the two
   MOS and one PN sensors for XMM-Newton.
   }\hss}}
  \end{tabular}
 \end{center}
\end{table*}

The XIS is equipped with four X-ray charge coupled devices (CCDs) placed at the focus of
four independent and co-aligned X-ray telescopes \citep{serlemitsos07}. The instrument
provides X-ray imaging-spectroscopic capability in the 0.2--12~keV band. Three
devices (XIS0, 2 and 3) are front-illuminated (FI), while the remaining one (XIS1) is
back-illuminated (BI). The FI and BI sensors have comparatively better response in
the hard and soft X-ray bands, respectively. The entire XIS2 and part of the XIS0 became
dysfunctional due to putative micro-meteorite hits since 2006 November and 2009 June,
respectively, and we used the remaining parts of the devices. The XIS covers an
18\arcmin$\times$18\arcmin\ field with a telescope half-power diameter of
$\sim$2\farcm0. Two radioactive $^{55}$Fe sources illuminate two corners of each device
for calibration purposes. The total effective area and the energy resolution in full
width at half maximum (FWHM) are 300~cm$^{2}$ and $\sim$ 170~eV at 6 keV for FI, and
250~cm$^{2}$ and $\sim$ 200~eV at 6 keV for BI, respectively. We operated the XIS in the
normal clocking mode with a frame time of 8~s in all observations. The relatively large
effective area and arguably the lowest and most stable instrumental background among all
previous X-ray CCD instruments in orbit give the XIS an advantage in
investigating diffuse X-ray emission with low surface brightness, such as the subject of
this study.

We retrieved the pipeline products and processed the data using the HEADAS software
package\footnote{See http://heasarc.gsfc.nasa.gov/lheasoft/ for detail.}  version 6.12
and the latest calibration database (CALDB) as of writing. We removed events taken
during passages over the South Atlantic Anomaly, those with elevation angles below 20 and 5 degrees
from the day and night Earth, respectively, and with the ASCA grades 1, 5, and 7. The net
exposure times are given in table~\ref{t1}.

\medskip

We also retrieved the archival XMM-Newton data \citep{nulsen11,panagoulia11} to
supplement the Suzaku data for investigating point-like X-ray sources that are potential
contamination for the diffuse emission of our interest. We found one data set
centered at the western lobe (figure~\ref{f1}). The XMM-Newton satellite
\citep{jansen01} carries three X-ray instruments. We used the data taken with the two
imaging-spectroscopic instruments: the European Photon Imaging Camera (EPIC) MOS
\citep{turner01} and PN \citep{strueder01}.

\section{Analysis \& Results}\label{s3}
\subsection{X-ray Images}\label{s3-1}
\begin{figure*}
 \begin{center}
  \FigureFile(185mm,85mm){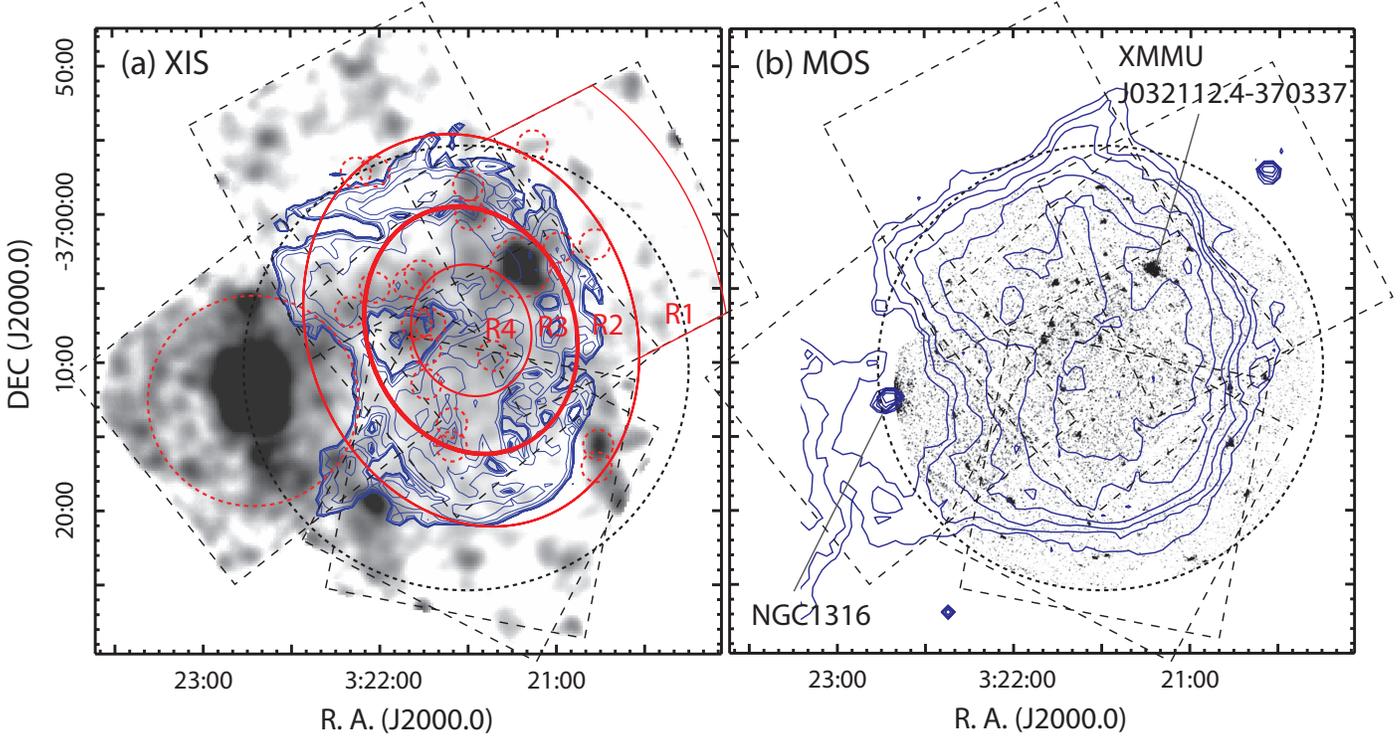}
 \end{center}
 \caption{(a) Suzaku XIS and (b) XMM-Newton MOS2 image of the western lobe and
 NGC\,1316. The XIS image is shown in the 0.67--1.5~keV band in logarithmic scale to
 emphasize the diffuse thermal emission, while the MOS image is shown in the 0.2--10~keV
 band in linear scale to highlight point-like sources. The blue contours indicate
 (a) linear polarized and (b) total intensities at 1.5~GHz \citep{fomalont89}. In (a),
 events from the three sensors were merged, and the differences in the exposure and the
 effective area were corrected, and the image was smoothed by a Gaussian convolution. The
 source extraction regions are shown as red curves; the thick ellipse (R3$+$R4) for
 characterizing the lobe spectrum, and the concentric elliptical annuli (R2--R4) as well as
 the flanking region (R1) for studying the spatial distribution. The masked circles around
 point-like sources and the central galaxy are shown as red dashed circles. In both images,
 the XIS and MOS fields of view are shown respectively as the dashed squares and
 the circle. The background was not subtracted in the images.}
 \label{f2}
\end{figure*}

Figure~\ref{f2} shows the (a) mosaicked XIS and (b) MOS images. The XIS image is shown
in the 0.67--1.5~keV band in logarithmic scale to focus on the diffuse emission from
thermal plasma with a temperature of $k_{\mathrm{B}}T\sim$1~keV as reported in previous
work \citep{feigelson95,kaneda95,tashiro01,tashiro09}, whereas the MOS image is in the
0.2--10~keV band in linear scale to highlight point-like sources.

Numerous point-like sources are recognized in the MOS image with the brightest one being
the nucleus of NGC\,1316. Another bright point-like source is found in the northwest quadrant of
the western lobe, which is an unidentified source (XMMU J032112.4--370337). Besides
these point-like sources, it is evident that diffuse emission exists in the XIS
image. The surface brightness distribution of the diffuse emission suggests that there
are two spatially distinct components: one is centered at NGC\,1316 (hereafter called
``the galaxy component''), and the other is in the western lobe (``the lobe
component''). To characterize the X-ray spectrum of the latter, we define the ``lobe
region'', shown with the thick red ellipse (R3$+$R4) in figure~\ref{f2} (a).

\subsection{X-ray Spectra (1) Background Emission}\label{s3-2}
As the emission of interest has a low surface brightness, we carefully subtracted
the background emission. We evaluated the contribution of (1) non X-ray background (NXB;
\S~\ref{s3-2-1}), (2) celestial X-ray diffuse background seen in all directions, and (3)
the possible intra-cluster gas (\S~\ref{s3-2-2}) by making a comparison with
other regions in the Fornax cluster.

For spectral fitting, we used the XSPEC package version 12.7. We generated the
redistribution matrix functions (RMF) using the \texttt{xisrmfgen} tool, and the
auxiliary response files (ARF) using the \texttt{xissimarfgen} tool
\citep{ishisaki07}. The two FI spectra with XIS0 and XIS3 were merged in view of their nearly
identical responses, while the BI spectrum was treated separately. When source spectra
were extracted using multiple data sets, the source and background spectra and the
response files were averaged, weighted by exposure times effective area using the
\texttt{addascaspec} tool. The fitting parameters were constrained to be the same between the FI and BI
spectra except for a normalization factor of a few percent to account for the internal
calibration uncertainties\footnote{See
http://www.astro.isas.jaxa.jp/suzaku/doc/suzakumemo/suzakumemo-2008-06.pdf for detail.}.

\subsubsection{Non X-ray background}\label{s3-2-1}
First, we considered the contribution from the NXB, which is caused mainly by charged
particles interacting with the sensors and surrounding structures, and is thus strongly
correlated with the geomagnetic cut-off-rigidity of the satellite orbit at the time of
observations. The level of the NXB of the XIS is so low and stable that it can be
reproduced by the accumulation of non-contemporaneous data taken while the telescope is
pointed toward the night Earth. Using the \texttt{xisnxbgen} tool, we simulated the NXB
spectrum by accumulating the night Earth data in such a manner that the cut-off-rigidity
histogram matches the actual observations. With this approach,
\citet{tawa08} demonstrated that the NXB spectrum is reproduced at an accuracy of
$\sim$5\% in the count rate.

For the FI spectra, we further fine-tuned the normalization of the simulated NXB
spectrum using instrumental features, most notably the Ni\emissiontype{I} K$\alpha$
emission line at 7.47~keV. We derived the intensity of the emission line in both the
actual and simulated spectra by phenomenologically fitting them with a power-law
continuum plus a Gaussian line model in the 6.7--8~keV band. The free parameters in the
model were the power-law normalization and index, and the Gaussian normalization. The
center energy and width of the Gaussian component were fixed to 7.47~keV and 0~eV,
respectively. We rescaled the simulated NXB spectra so that the Ni\emissiontype{I}
K$\alpha$ line intensity became equal to that of the actual data.

For the BI spectra, the approach for the FI cannot be taken as the NXB level is much
higher in the high energy end of the spectrum including the Ni\emissiontype{I} K$\alpha$
line. Instead, we used the count rate in the 8--10~keV range to derive the rescaling factor,
where the NXB emission dominates over the celestial X-ray emission.

\subsubsection{Celestial X-ray diffuse background}\label{s3-2-2}
For the celestial background emission seen in any direction of the sky, two sources make
major contributions in the energy range and the surface brightness that we study: (1)
the Galactic halo emission, which is characterized by optically-thin thermal plasma
emission with a temperature of 0.2--0.4~keV and a surface brightness of $\approx
10^{-8}$~erg~s$^{-1}$~cm$^{-2}$~str$^{-1}$ \citep{lumb02} and (2) the cosmic X-ray
background (CXB) emission, which is characterized by a power-law spectrum with a photon
index of $\sim$1.4 and a surface brightness of $\approx
10^{-7.5}$~erg~s$^{-1}$~cm$^{-2}$~str$^{-1}$ \citep{kushino02}. In addition, there
should be diffuse emission pervasive in the Fornax cluster that NGC\,1316 belongs to.

We evaluated the contribution of these components by making a comparison with other XIS
data in near-by fields. For~A is displaced from the cluster center by 3.6\arcdeg, and we
retrieved the archival data taken at similar cluster-centric distances. Three data
sets devoid of bright point-like sources were found (table~\ref{t2}). All the data were
taken with the normal clocking mode.

\begin{table*}
 \begin{center}
  \caption{Suzaku data sets for the comparison regions in the Fornax cluster.}
  \label{t2}
  \begin{tabular}{cccccc}
   \hline
   \hline
   Sequence ID & \multicolumn{2}{c}{Position (J2000.0)}    & Observation & Dist.\footnotemark[$*$] & $t_{\mathrm{exp}}$\footnotemark[$\dagger$]\\
               & R.\,A.  & DEC                             & start date  & (arcdeg) &  (ks)$^{b}$\\
   \hline
   703038010 & \timeform{03h31m06.3s} & \timeform{-38D24'05''} & 2008-06-16 & 3.30 & 24.3\\
   802037010 & \timeform{03h13m10.4s} & \timeform{-37D40'25''} & 2007-06-28 & 5.55 & 15.0\\
   802040010 & \timeform{03h19m57.4s} & \timeform{-32D03'58''} & 2007-06-29 & 5.13 & 19.7\\
   \hline
   \multicolumn{6}{@{}l@{}}{\hbox to 0pt{\parbox{130mm}{\footnotesize
   \par\noindent
   \footnotemark[$*$] The distance from the center of the Fornax cluster at (R.\,A.,
   Decl.)$=$(\timeform{03h38m30.9s}, \timeform{-35D27'16''}).
   \par\noindent
   \footnotemark[$\dagger$] The mean exposure time of the three XIS sensors.
   }\hss}}
  \end{tabular}
 \end{center}
\end{table*}

We identified point-like sources in the XIS images (0.5--5.5~keV) for the For~A region and the
three comparison regions using the sliding cell technique in the \texttt{XIMAGE}
package. We masked a circle of 1\arcmin\ in radius around these sources. In addition,
for a few bright sources including XMMU J032112.4--370337, we expanded the masking
radius up to 2\arcmin\ so that the events spilled outside of the masks are negligible in
comparison to the background events. We also masked the region within 7\arcmin\ around
NGC\,1316. The masked regions are shown as the red dashed circles in figure~\ref{f2}
(a) for the For~A region.

We fitted the NXB-subtracted spectra with a model consisting of the Galactic halo and
the CXB components. For the Galactic halo component, we used the \texttt{APEC} model
\citep{smith01} with the plasma temperature ($k_{\mathrm{B}}T^{\mathrm{(GH)}}$) and the
surface brightness ($S^{\mathrm{(GH)}}$) as free parameters and the metal abundance
fixed to that of solar \citep{anders89}. For the CXB component, we used the power-law
model with the surface brightness ($S^{\mathrm{(CXB)}}$) as a free parameter and the
power-law index fixed to 1.4 \citep{kushino02}. These components were attenuated by an
interstellar photo-electric absorption model with an absorption column fixed to the
Galactic value toward each region \citep{kalberla05}.

\begin{figure*}
 \begin{center}
  \FigureFile(180mm,180mm){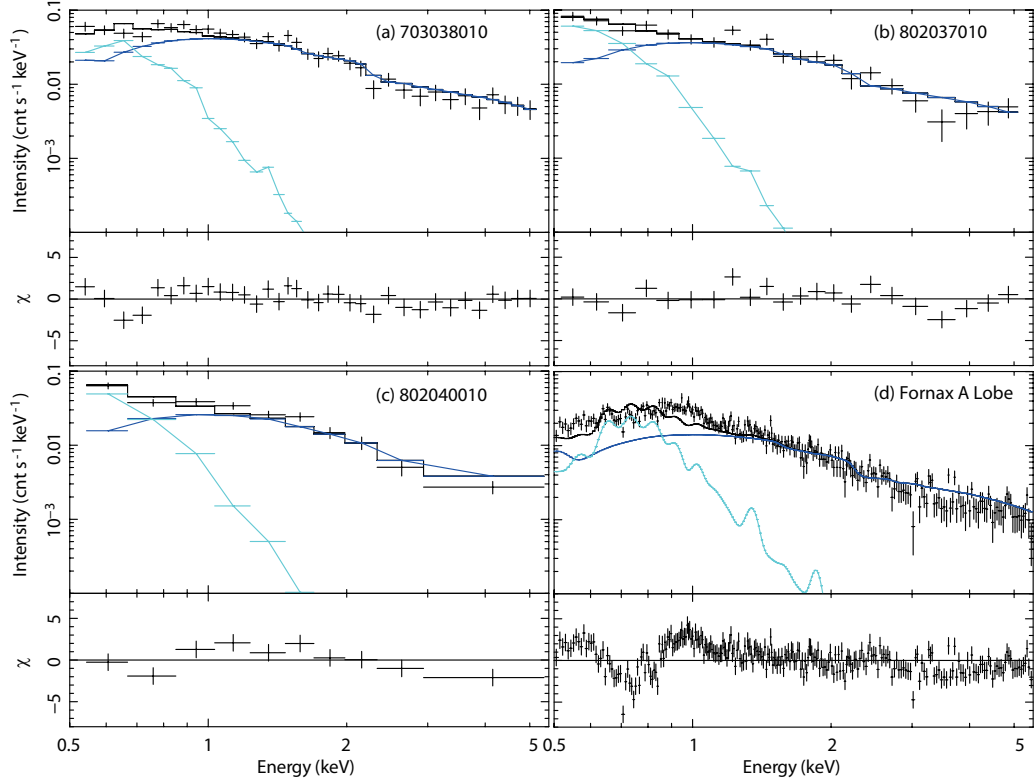}
 \end{center}
 \caption{Spectra and the best-fit models of the diffuse spectrum of (a)--(c) the three
 comparison regions and (d) the lobe region. The spectra were binned with
 100~counts~bin$^{-1}$ except for (c) with 150~counts~bin$^{-1}$ to compensate for its
 poorer statistics. The model is composed of the Galactic halo (cyan) and CXB (blue)
 components. For (d), the temperature of the thermal emission was fixed to
 $k_{\mathrm{B}}T=$0.2~keV to reproduce the Galactic halo emission, rather than the lobe
 thermal emission.}
 \label{f3}
\end{figure*}

Figure~\ref{f3} shows the spectra and the best-fit models, while table~\ref{t3}
summarizes the best-fit parameters. The spectra in the three comparison regions were
well fitted by the Galactic halo plus CXB model. All the parameters are consistent with
the previous studies in different regions \citep{lumb02,kushino02,yoshino09}. No
additional component was required, suggesting that the diffuse thermal emission in the
Fornax cluster, if any, does not make a significant contribution at such large
cluster-centric distances. This is consistent with the ROSAT result \citep{jones97}, 
in which the radial profile of the surface brightness of the Fornax cluster gas was
presented out to 280~kpc ($\sim$40\arcmin) from the cluster center. The observed value at
200~kpc is $1 \times 10^{-8}$~erg~s$^{-1}$~cm$^{-2}$~str$^{-1}$ and the extrapolated
value at the position of Fornax A is $<$10$^{-10}$~erg~s$^{-1}$~cm$^{-2}$~str$^{-1}$,
which is far below our detection level. 

In contrast, the same model was found inadequate for the spectrum in the lobe
region. The largest difference from the other three regions is the excess emission
peaking at $\sim$1~keV, indicating the presence of extra emission local to the lobe
region as we noticed in figure~\ref{f2} (a).

\begin{table*}
 \begin{center}
  \caption{Fitting results for the comparison regions\footnotemark[$*\dagger$].}
  \label{t3}
  \begin{tabular}{lccccc}
   \hline
   \hline
   Region & 
   \multicolumn{1}{c}{Absorption} &
   \multicolumn{2}{c}{------ Galactic halo ------} &
   \multicolumn{1}{c}{------ CXB ------} &
   \multicolumn{1}{c}{$\chi^{2}$/d.o.f.} \\
   &
   $N_{\mathrm{H}} / 10^{20}$ & 
   $k_{\mathrm{B}}T^{\mathrm{(GH)}}$ &
   $S^{\mathrm{(GH)}} / 10^{-8}$ &
   $S^{\mathrm{(CXB)}} / 10^{-8}$ &
   \\
   &
   (1) &
   (2) &
   (3) &
   (4) &
   (5) 
   \\
   \hline
   703038010 & 1.33 & 0.20 $^{+0.04}_{-0.01}$ & 1.04 $^{+0.13}_{-0.13}$ & 7.92 $^{+0.25}_{-0.25}$ &                         101.4/71\\
   802037010 & 1.77 & 0.18 $^{+0.01}_{-0.01}$ & 1.42 $^{+0.17}_{-0.16}$ & 6.73 $^{+0.21}_{-0.23}$ &                          63.4/41\\
   802040010 & 1.34 & 0.19 $^{+0.01}_{-0.01}$ & 1.19 $^{+0.15}_{-0.14}$ & 4.94 $^{+0.22}_{-0.22}$ &                          33.6/22\\
   \hline
   \multicolumn{6}{@{}l@{}}{\hbox to 0pt{\parbox{130mm}{\footnotesize
   \par\noindent
   \footnotemark[$*$] The best-fit parameters for 
   (1) interstellar photo-electric absorption $N_{\mathrm{H}}$ (cm$^{-2}$), 
   (2) plasma temperature $k_{\mathrm{B}}T^{\mathrm{(GH)}}$ (keV),
   (3) 0.5--2.0~keV surface brightness $S^{\mathrm{(GH)}}$ (erg~s$^{-1}$~cm$^{-2}$~str$^{-1}$) of the Galactic halo, 
   (4) 2.0--10.0~keV surface brightness $S^{\mathrm{(CXB)}}$ (erg~s$^{-1}$~cm$^{-2}$~str$^{-1}$) of the CXB, 
   (5) $\chi^{2}$ value and the degree of freedom.
   \par\noindent
   \footnotemark[$\dagger$] The error ranges indicate 1 $\sigma$ statistical uncertainty. 
   The values without errors are those fixed during the fitting.
   \par\noindent
   }\hss}}
  \end{tabular}
 \end{center}
\end{table*}

\subsection{X-ray Spectra (2) Lobe Emission}\label{s3-3}
\subsubsection{Thermal emission}\label{s3-3-1}
\begin{figure}
 \begin{center}
  \FigureFile(85mm,85mm){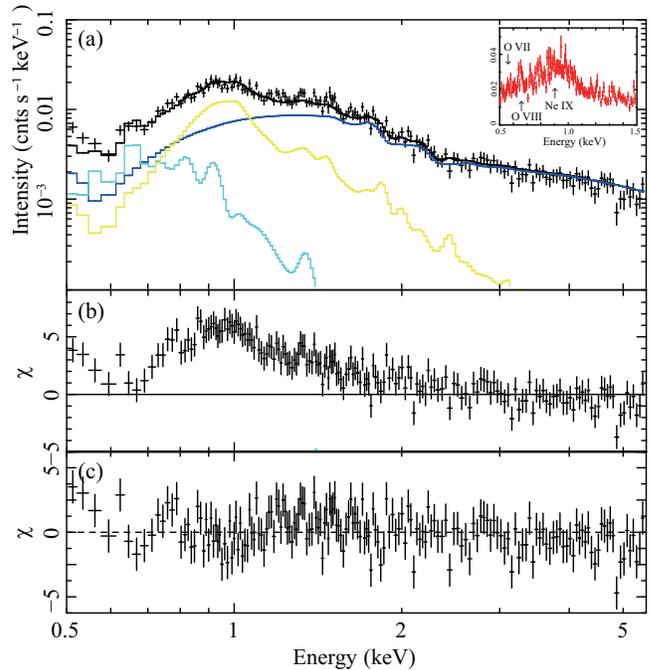}
 \end{center}
 \caption{Spectral fitting of the lobe emission. Only the FI results are shown for
 clarity unless otherwise noted. (top) The NXB-subtracted spectra and the best-fit
 spectral model. Each component is color-coded: yellow for the lobe thermal, cyan for
 the Galactic halo, and blue for the CXB. A close-up view of the BI spectrum in the soft
 band is given in the inset. (middle) The residuals to the fit by the model consisting
 only of the Galactic halo and the CXB components. (bottom) The residuals to the fit by
 the model consisting of the lobe thermal, Galactic halo, and the CXB components.}
 \label{f4}
\end{figure}

In a close-up view of the lobe spectrum in the soft-band (inset in the top panel in
figure~\ref{f4}), several emission features are recognized. While the
O\emissiontype{VII} and O\emissiontype{VIII} lines respectively at 0.57 and 0.65~keV are
from the Galactic halo component with a temperature of $k_{\mathrm{B}}T=$0.2~keV, the
Ne\emissiontype{IX} line at 0.91~keV and Fe L series lines at $\sim$1~keV should be from
a thermal plasma of a higher temperature.

Therefore, we added a thermal plasma component represented by the \texttt{APEC} model
with free parameters for the plasma temperature ($k_{\mathrm{B}}T^{\mathrm{(lobe,T)}}$)
and the surface brightness ($S^{\mathrm{(lobe,T)}}$) and a fixed value of 0.3 solar for the abundance.
The residuals were reduced substantially (the middle and the bottom
panels in figure~\ref{f4} respectively show the results without and with the additional
thermal component) and yielded a statistically acceptable result. The best-fit result is
summarized in table~\ref{t5}.

\begin{table*}
 \begin{center}
  \caption{Fitting results for the lobe spectrum.}
  \label{t5}
  \begin{tabular}{lcccccc}
   \hline
   \hline
   &
   \multicolumn{2}{c}{--- Galactic halo ---} &
   \multicolumn{1}{c}{--- CXB ---} &
   \multicolumn{2}{c}{--- Thermal ---} &
   \multicolumn{1}{c}{$\chi^{2}$/d.o.f.} \\
   &
   $k_{\mathrm{B}}T^{\mathrm{(GH)}}$ &
   $S^{\mathrm{(GH)}} / 10^{-8}$ &
   $S^{\mathrm{(CXB)}} / 10^{-8}$ &
   $k_{\mathrm{B}}T^{\mathrm{(lobe,T)}}$ &
   $S^{\mathrm{(lobe,T)}} / 10^{-8}$\footnotemark[$\ddagger$] &
   \\
   &
   (keV) &
   \multicolumn{2}{c}{(erg~s$^{-1}$~cm$^{-2}$~str$^{-1}$)} &
   (keV) &
   (erg~s$^{-1}$~cm$^{-2}$~str$^{-1}$) &
   \\
      &
  (1) &
  (2) &
  (3) &
  (4) &
  (5) &
  (6) \\
   \hline
   Without NT & 0.20 $^{+0.01}_{-0.01}$ & 1.36 $^{+0.07}_{-0.07}$  & 8.17 $^{+0.12}_{-0.12}$ & 0.99 $^{+0.01}_{-0.01}$ & 1.90 $^{+0.06}_{-0.06}$ & 426.4/319\\
   With NT    & 0.19 $^{+0.01}_{-0.01}$ & 1.28 $^{+0.07}_{-0.07}$  & 5.85 $^{+0.12}_{-0.12}$ & 0.98 $^{+0.02}_{-0.02}$ & 1.78 $^{+0.06}_{-0.06}$ & 395.9/319\\
   \hline
   \multicolumn{7}{@{}l@{}}{\hbox to 0pt{\parbox{170mm}{\footnotesize
   \par\noindent
   \footnotemark[$*$] The best-fit parameters for the fitting with and without the lobe
   non-thermal (NT) component in the model. The parameters are
   (1) plasma temperature 
   (2) 0.5--2.0~keV surface brightness of the Galactic halo,
   (3) 2.0--10.0~keV surface brightness of the CXB,
   (4) plasma temperature,
   (5) 0.5--2.0~keV surface brightness of the lobe thermal emission, and
   (6) $\chi^{2}$ value and the degree of freedom.
   The photo-electric absorption column density was fixed to 2.06 $\times$ 10$^{20}$~cm$^{-2}$, while
   the power-law index and the surface brightness of the non-thermal component was fixed
   to 1.68 and 2.0$\times$10$^{-8}$~erg~s$^{-1}$~cm$^{-2}$~str$^{-1}$ in the
   2.0--10.0~keV band.
   \par\noindent
   \footnotemark[$\dagger$] The error ranges indicate 1 $\sigma$ statistical uncertainty. 
   \par\noindent
   \footnotemark[$\ddagger$] A uniform distribution over a 12\arcmin\ radius disk was assumed.
   }\hss}}
  \end{tabular}
 \end{center}
\end{table*}

\subsubsection{Contaminating emission}\label{s3-3-2}
We next evaluated how these parameters are affected by contaminating emission that are
unseen in the comparison regions: the point-like sources (\S~\ref{s3-3-2-1}), the
non-thermal emission in the lobe (\S~\ref{s3-3-2-2}), and 
the diffuse emission associated with the central galaxy NGC\,1316 (\S~\ref{s3-3-2-3}).

\paragraph{Point-like sources}\label{s3-3-2-1}
In order to see the contribution by point-like sources, we utilized the XMM-Newton data,
which has much better X-ray optics, hence is more sensitive to point-like sources. We
started with the XMM-Newton source list in the pipeline products with a detection
significance above 5$\sigma$. For each individual
source, we accumulated source and background events from a 36\arcsec\ radius circle and
a 72--180\arcsec\ annulus, respectively. In the background annulus, 36\arcsec\ radius
circles around other sources were masked. The RMF and ARF were generated using the
\texttt{rmfgen} and \texttt{arfgen} tools.

We constructed the composite spectrum of all the XMM-Newton point-like sources except
for XMMU J032112.4--370337, and fitted it successfully with an absorbed power-law model
with the best-fit photon index of 2.22$^{+0.04}_{-0.03}$ and the surface brightness of
2.57 $^{+0.16}_{-0.15}$ $\times$ 10$^{-8}$~erg~s$^{-1}$~cm$^{-2}$~str$^{-1}$ in
2.0--10.0~keV band, assuming that the summed flux is distributed uniformly across the lobe
region. We do not include this component in the model to fit the XIS spectrum to avoid a
duplicated count with the CXB emission, which is mostly the ensemble of such point-like
sources of background AGNs \citep{giacconi01}. We also confirmed that there is no local
excess of point-like source populations in the lobe region by comparing with the
XMM-Newton source list constructed in one of the comparison regions (sequence ID
703038010 in table~\ref{t2}). A somewhat softer spectrum of the composite point-like
sources in comparison with the CXB may be due to contamination by the diffuse thermal emission
into the apertures around each source. Anyway, the surface brightness of the point-like
sources is only about 1/10 of that of the lobe thermal emission at 1~keV.  We thus
conclude that the ensemble of point-like sources does not account for the observed
excess emission in the lobe region.

\paragraph{Non-thermal emission}\label{s3-3-2-2}
Following the previous work \citep{tashiro09}, we further added the lobe non-thermal
component originating from inverse Compton emission to the model. The photon index
and the surface brightness were fixed to 1.68 and
2.0$\times$10$^{-9}$~erg~s$^{-1}$~cm$^{-2}$~str$^{-1}$ in the 2.0--10.0~keV band
\citep{tashiro09}. The fitting improved slightly with little changes in the best-fit
parameters of the lobe thermal component (table~\ref{t5}). We thus conclude that the
lobe thermal emission is present regardless of whether or not we include the lobe non-thermal
component in the model.

\paragraph{Galaxy component}\label{s3-3-2-3}
\citet{kim98} and \citet{kim03} revealed, using the high-resolution X-ray imagers
on-board the ROSAT and Chandra X-ray Observatory respectively, that diffuse X-ray
emission exists around NGC\,1316 in smoothed X-ray images after removing point-like
sources. The morphology of the diffuse emission is elongated in the direction normal to
the jet launching direction at its root. The extent of the diffuse emission is quite
limited; the brightness drops by an order of magnitude in 10\arcsec\ in the radial profile (see
figure 6 in \cite{kim03}). The spectrum was modeled with a thermal plasma with a
temperature of $k_{\mathrm{B}}T=$0.5--0.6~keV.

\citet{konami10} used the same XIS data presented here (sequence ID 801015010 in
figure~\ref{f1}) to analyze the X-ray spectrum of the diffuse emission in
NGC\,1316. Despite the limited imaging capability of Suzaku, they identified the
spectrum of the diffuse plasma emission with its most prominent features of
Mg\emissiontype{XI} and Si\emissiontype{XIII} emission lines. These features are absent
in their background region taken from 5\arcmin\ away from the center of the galaxy, for
which the spectrum was explained only by background emission. The lobe region is
separated from NGC\,1316 by $\sim$ 8\arcmin\, so we can safely assume that our diffuse
X-ray spectrum in the lobe is free from contamination by the galaxy component.

\subsection{Spatial Extent of the Thermal Emission}\label{s3-4}
\begin{figure}
 \begin{center}
  \FigureFile(85mm,85mm){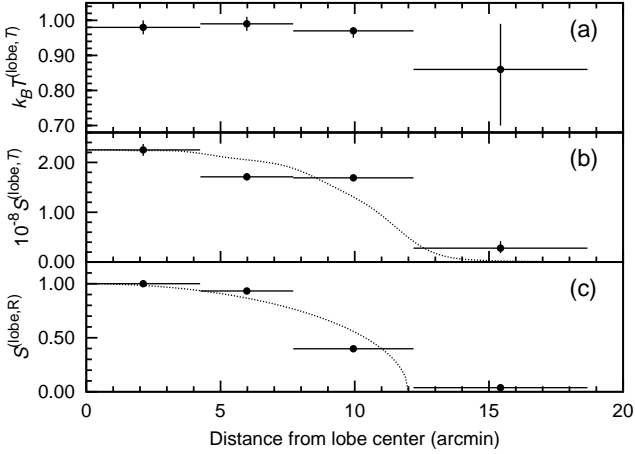}
 \end{center}
 \caption{Radial profiles of (a) the plasma temperature
 ($k_{\mathrm{B}}T^{\mathrm{(lobe,T)}}$) and (b) the surface brightness
 $S^{\mathrm{(lobe,T)}}$ of the diffuse thermal emission, and (c) the 1.5~GHz total
 intensity \citep{fomalont89} from the center of the western lobe. The horizontal error
 bars are the region size calculated as the logarithmic mean of the major and minor axes
 of the ellipses (R1--R4; figure~\ref{f2}b), while the vertical error bars correspond to 1
 $\sigma$ statistical uncertainty.  The curves in (b) and (c) show the fully-resolved (c)
 and PSF-convolved (b) radial profiles of completely optically-thin emission from a uniformly filled
 sphere of 12\arcmin\ radius.}
 \label{f5}
\end{figure}

Finally, we investigated the spatial distribution of the lobe thermal emission. We
divided the regions inside and outside of the lobe radially into three elliptical annuli
(R2--R4; figure~\ref{f2}b). Here, R3$+$R4 is the lobe region used for the analyses in
the previous sections. We also defined the R1 region as a concentric ellipse outside of
the R2 region and within the field of the sequence ID $=$ 804036010 (figure~\ref{f1}),
which is intended to examine the local background unaffected by the central galaxy and the
eastern lobe.  For each region, we conducted spectral modeling in the same approach
as in \S~\ref{s3-3-1}. The results are shown in figure~\ref{f5}.

We compared the observed radial profile with the projected profile calculated by
assuming that the thermal gas is optically thin and uniformly fills a spherical volume with a
radius of 12\arcmin\ (the approximate size of the western lobe as seen in the 1.5 GHz radio band;
figure~\ref{f2}) around the center of the western lobe at (RA, DEC) $=$
(\timeform{03h21m29s}, \timeform{-37D07'52''}) in the equinox J2000.0. The dashed curve
in figure~\ref{f5} (c) shows the case when the spatial distribution is fully resolved,
while that in figure~\ref{f5} (b) shows the case when the distribution is convolved with
the point spread function of the XIS. The observed profile of the thermal X-ray emission
is consistent with the assumed distribution, and there is no indication of
relative enhancement toward the outer edges of the lobe that would suggest a shell-like structure.

\section{Discussion}\label{s4}
Our Suzaku observations have confirmed the presence of the diffuse thermal emission in
the western lobe of For~A. We argued that this emission is local to the lobe, unlike
those from the Galactic halo, CXB, and Fornax cluster gas, because it is not found in
the three comparison regions (\S~\ref{s3-2}). We also argued that the emission cannot be
explained by the ensemble of point-like sources, the non-thermal emission in the lobe,
nor the contaminating hot gas from the NGC\,1316 galaxy (\S~\ref{s3-3}). The surface
brightness distribution is consistent with optically-thin plasma emission of a uniform
spherical distribution whose position and extent are comparable to those observed in the
radio band (\S~\ref{s3-4}).

\medskip

In view of evidence for thermal plasma residing within the lobe, we
estimate its physical attributes such as the pressure and energy balance,
as well the total mass and energy. The best-fit \texttt{APEC} normalization
$N_{\mathrm{APEC}}$ is related to the electron density $n_{\mathrm{e}}$ as
\begin{equation}
 4 \pi D^{2}(1+z)^{2} \times 10^{14} N_{\mathrm{APEC}} = n_{\mathrm{e}}^2 Vf,
\end{equation}
where $D$ is the distance to For A (18.6~Mpc), $z$ is the redshift (4.65 $\times$
10$^{-3}$), $V$ is the western lobe volume, and $f$ is the volume filling factor.  We
assume for simplicity that the plasma consists only of ionized H and electrons. For the
emitting volume $fV$, we consider a uniform sphere with an angular radius of
$\theta=$12\arcmin\ (figure~\ref{f5}), so that $V=4\pi (D\theta)^{3}/3$. Using the
best-fit values from spectral fitting in table~\ref{t5}, we derive the electron density
$n_{\mathrm{e}}=3.0 \times 10^{-4} f^{-1/2}$~cm$^{-3}$, the thermal gas pressure
$p_{\mathrm{T}}=2n_{\mathrm{e}}k_{\mathrm{B}}T f^{-1/2} = 1\times 10^{-12} f^{-1/2}
$~erg~cm$^{-3}$, the total thermal energy
$E_{\mathrm{T}}=3n_{\mathrm{e}}k_{\mathrm{B}}TfV \sim 5 \times 10^{58} f^{1/2}$~erg, and
the total gas mass $M_{\mathrm{T}}=m_{\mathrm{p}} n_{\mathrm{e}}fV \sim 9 \times 10^{9}
f^{1/2}$\,$M_{\odot}$, where $m_{\mathrm{p}}$ is the proton mass.
The filling factor is constrained to be $f^{1/2} > 10^{-3}$, as the radiative cooling
time of the gas $\tau_{\mathrm{r,c}} \sim E_{\mathrm{T}}/L_{\mathrm{T}} \approx 10^{3}
f^{1/2}$~Gyr must not be shorter than the radio lobe's age of 0.1~Gyr \citep{iyomoto98},
where $L_{\mathrm{T}} \sim S^{\mathrm{(lobe,T)}} \left(D(1+z)\theta\right)^{2}$ is the
thermal X-ray luminosity.

A comparison of the energy densities in the thermal gas $\epsilon_{\mathrm{T}}$, in the
magnetic field $\epsilon_{\mathrm{mag}}$ and in non-thermal electrons
$\epsilon_{\mathrm{NT,e}}$ is shown in table~\ref{t6}. \citet{tashiro09} derived
$\epsilon_{\mathrm{mag}} = 6.7 \times 10^{-14}$~erg~cm$^{-3}$ and
$\epsilon_{\mathrm{NT,e}}=5.0 \times 10^{-13}$~erg~cm$^{-3}$ under the assumption that
the observed radio and X-ray emission are respectively due to synchrotron radiation in a
homogeneous magnetic field and inverse Compton upscattering of cosmic microwave
background photons.  If the volume filling factor is unity for all components,
$\epsilon_{\mathrm{T}}/\epsilon_{\mathrm{NT,e}} \sim$3, so the thermal plasma is not far
from energy equipartition with the non-thermal electrons, even though
$\epsilon_{\mathrm{mag}}$ is about an order of magnitude smaller than either of the
other two.

As mentioned in \S~\ref{s3-2-2}, the thermal emission from the ICM of the Fornax cluster
is not directly observable in the vicinity of For A.  In order to examine the pressure
balance between the For A lobe and the ambient ICM, we must rely on extrapolations from
the properties of the ICM observed in the inner regions by \citet{jones97}.  For the
electron density, the beta-model profile determined at cluster radii 35--280~kpc is
extrapolated to the position of For A at $\sim$1500~kpc.  Although the temperature
profile was seen to be declining with radius, the value of the temperature itself was
not well constrained, so we use the outermost observed value of 1.1~keV at 120~kpc as an
upper limit at the position of For A.  In this manner, the ICM pressure around the For A
lobe is estimated to be $\lesssim$ 2 $\times$ 10$^{-13}$~erg~cm$^{-3}$.  The total
pressure within the lobe due to the thermal gas, non-thermal electrons and magnetic
field is larger than this value by a factor of $\gtrsim$5, implying that the lobe is
over-pressured with respect to the ambient medium, although not by a very large
amount.

The above properties of the thermal gas in the For A lobe can be compared with those
tentatively identified recently in the giant lobes of the nearby radio galaxy Cen A
through two lines of evidence; diffuse thermal X-ray emission, albeit in a small portion
of the entire lobe \citep{stawarz13}, and excess Faraday rotation measures in radio
polarization maps \citep{osullivan13}. In contrast to our case here for For~A, the two
Suzaku pointings by \citet{stawarz13} only cover a limited area of Cen A's southern
giant lobe, and the spectra and spatial distribution of the thermal X-ray emission
across the lobe is unavailable.  Although the polarization maps of \citet{osullivan13}
cover the whole lobe structure, the measured rotation measures are highly non-uniform
and their interpretation is complicated by the unknown structure of the magnetic field.
Nevertheless, \citet{stawarz13} and \citet{osullivan13} estimated volume-averaged
quantities for the thermal gas that are consistent with each other, which are listed in
table~\ref{t6}.  For the Cen A lobe, $\epsilon_{\mathrm{T}}$, $\epsilon_{\mathrm{mag}}$
and $\epsilon_{\mathrm{NT,e}}$ all seem to be roughly in equipartition with each other.
Although this is unlike For A for which $\epsilon_{\mathrm{mag}} \ll
\epsilon_{\mathrm{NT,e}}$, the ratio of energies in the thermal gas relative to the sum
of those in the non-thermal components
$\epsilon_{\mathrm{T}}$/($\epsilon_{\mathrm{mag}}+\epsilon_{\mathrm{NT,e}}$)$\sim$3
appears to be similar for both objects.  Assuming such gas to be distributed throughout
the entire lobe of Cen A, the total estimated mass of $M_{\mathrm{T}}
\sim10^{10}$\,$M_{\odot}$ \citep{osullivan13} is also similar to our value for For A.

\begin{table*}
 \begin{center}
  \caption{Physical parameters of the For~A and Cen A lobes.}
  \label{t6}
  \begin{tabular}{lccccccc}
   \hline
   \hline
   Object &
   $n_{\mathrm{e}}$ &
   $k_{\mathrm{B}}T$ &
   $V$ &
   $\epsilon_{\mathrm{T}}$ &
   $\epsilon_{\mathrm{mag}}$ &
   $\epsilon_{\mathrm{NT,e}}$ &
   $R$
   \\
   &
   (cm$^{-3}$) &
   (keV) &
   (cm$^{3}$) &
   (erg~cm$^{-3}$) &
   (erg~cm$^{-3}$) &
   (erg~cm$^{-3}$) &
   \\
   &
  (1) &
  (2) &
  (3) &
  (4) &
  (5) &
  (6) &
  (7) \\
   \hline
   For A$\dagger$  & $3.0 \times 10^{-4}$ & 1.0 & $3.4 \times 10^{70}$ & $1.4 \times 10^{-12}$ & $6.7 \times 10^{-14}$ & $5.0 \times 10^{-13}$ & 2.5\\
   Cen A$\ddagger$ & 0.9--2.5$\times 10^{-4}$ & 0.5 & $2.0 \times 10^{71}$ & $2.4 \times 10^{-13}$ & $4.0 \times 10^{-14}$ & $5.2 \times 10^{-14}$ & 2.6\\
   \hline
   \multicolumn{8}{@{}l@{}}{\hbox to 0pt{\parbox{170mm}{\footnotesize
   \par\noindent
   \footnotemark[$*$] 
   The parameters are (1) electron density, (2) temperature, and (3) emitting volume
   (cm$^3$) assuming a volume filling factor of unity for the lobe thermal
   emission. The energy densities of (4) the thermal emission, (5) magnetic field, and
   (6) non-thermal electrons. (7) The ratio of energies in thermal to non-thermal components
   defined as
   $\epsilon_{\mathrm{T}}$/($\epsilon_{\mathrm{mag}}+\epsilon_{\mathrm{NT,e}}$).
   \par\noindent
   \footnotemark[$\dagger$] The values are from this work and \citet{tashiro09}.
   \par\noindent
   \footnotemark[$\ddagger$] The values are from \citet{stawarz13,abdo10}.
   }\hss}}
  \end{tabular}
 \end{center}
\end{table*}

\medskip

Finally, we speculate on the origin of the thermal gas inside the For A lobe.
Its total mass of $M_{\mathrm{T}} \sim 9 \times 10^{9} f^{1/2} M_{\odot}$
can be compared with the mass of the central black hole
$M_{\mathrm{BH}}\sim (1-2) \times 10^{8}$\,$M_{\odot}$ \citep{kuntschner00},
and the total gas mass currently in the host galaxy $M_{\mathrm{g,host}} \sim 3.3 \times 10^{9}$\,$M_{\odot}$ \citep{forman85}.
Assuming a filling factor of unity, $M_{\mathrm{T}} \sim 100 \times M_{\mathrm{BH}}$,
which immediately disfavors the possibility that it was transported outward from the nucleus inside the jets,
since it is highly improbable that 100 times more mass can be ejected than it can be accreted onto the black hole.

Therefore, the gas must have somehow been transported into the lobe from its surroundings on larger scales.
The fact that $M_{\mathrm{T}} \gg M_{\mathrm{BH}}$ also entails that this transport process
must inevitably be accompanied by strong deceleration of the outflow.
The jet kinetic energy can be written $E_{\rm jet}=\eta_{\rm jet} M_{\rm acc} c^2$,
where $M_{\rm acc}$ is the total mass accreted onto the black hole,
and $\eta_{\rm jet}$ is the corresponding efficiency of conversion from rest mass energy.
If the jet is loaded with baryonic material with mass $M_{\rm jet}$ from its environs at some location,
its bulk Lorentz factor $\Gamma_{\rm jet}=1 + E_{\rm jet}/M_{\rm jet} c^2 \le 1+ \eta_{\rm jet} M_{\rm BH}/M_{\rm jet}$,
as $M_{\rm acc} \le M_{\rm BH}$.
Considering physically plausible values of $\eta_{\rm jet} < 1$,
if $M_{\rm jet} \sim M_{\rm BH}$, $\Gamma_{\rm jet} \lesssim 2$ and the jet must be decelerated to sub-relativistic velocities,
as is often invoked to explain the morphological properties of FR I radio galaxies \citep{bicknell95}.
If $M_{\rm jet} \gg M_{\rm BH}$, $\Gamma_{\rm jet} \rightarrow 1$,
and the deceleration is likely to be so severe as to disrupt the robust development of jet structure.
Hence, gas with mass as large as $M_{\mathrm{T}}$
cannot have been entrained into the jet during its passage through the interstellar medium (ISM) of the galaxy,
as it would preclude the formation of well-defined lobes on much larger scales as observed,
even though $M_{\mathrm{g,host}}$ is seen to be comparable to $M_{\mathrm{T}}$.

Thus, the gas is likely to have been entrained into the lobe after it has sufficiently
expanded and evolved, perhaps after cessation of the jet energy supply, through
e.g. Rayleigh-Taylor instabilities in the interface between the lobe and the external
medium \citep{reynolds02}.  The material itself may be either the ambient ICM, or ISM
that has been swept up and accumulated outside the lobe (but not entrained) during
propagation in the host galaxy.  In either case, the implication is that the radio jets
and lobes have exerted a profound influence on the gaseous medium of its host galaxy
and/or cluster.  Further discussion on the origin of the thermal gas must await more
detailed observations of this and other objects.

\section{Summary}\label{s5}
We presented the results of X-ray mapping observations of the western radio lobe of the
For~A galaxy using the Suzaku XIS. Comparing with the Suzaku data in three near-by
regions and XMM-Newton data in the lobe, we concluded that diffuse thermal emission with
temperature $k_{\mathrm{B}}T \sim$1~keV is present in the western part of the lobe in
excess of background and contaminating emission. The surface brightness distribution is
consistent with a uniform spherical distribution centered within the western lobe with a
projected radius of 12\arcmin.

We derived the physical parameters of the thermal gas responsible for the diffuse
emission.  Its total mass and energy amounts respectively to $\sim 9 \times 10^{9}
f^{1/2}$\,$M_{\odot}$ and $\sim 5 \times 10^{58} f^{1/2}$~erg.  Such a large mass
disfavors the possibility that the gas was ejected from the central engine, and instead
implies that it was transported into the jet from its surroundings on large scales, a
potential indication of significant interaction and feedback of the radio jet with the
host galaxy and/or cluster.  Some properties, such as the approximate equipartition of
the thermal and non-thermal components and the total gas mass appear to be similar to
the thermal gas tentatively found in the giant lobe of the Cen A radio galaxy.  Future,
more detailed studies, particularly of the spatial distribution of such emission
components in For A and other radio galaxies are warranted to elucidate its origin and
its implications.

\bigskip

We acknowledge N. Isobe for useful discussion. H.S. is supported by the Research Center
for Measurement in Advanced Science in Rikkyo University. MST and SI are supported in
part by the Grants-in-Aid for Scientific Research Nos. 22340039 and 22540278 from MEXT
of Japan, respectively. This research made use of data obtained from Data ARchives and
Transmission System (DARTS), provided by Center for Science-satellite Operation and Data
Archives (C-SODA) at ISAS/JAXA.

\bibliographystyle{aa}
\bibliography{ms}

\end{document}